\documentclass[amsmath,amssymb,prl,showpacs,twocolumn]{revtex4}

\usepackage{graphicx}
\usepackage{times}

\newcommand{\bra}[1]{\langle #1|}
\newcommand{\ket}[1]{| #1 \rangle}

\newcommand{\cre}[1]{\hat{#1}^\dagger}
\newcommand{\des}[1]{\hat{#1}}

\begin{document}

\title{Single-photon Optomechanics}

\author{A.~Nunnenkamp}
\affiliation{Departments of Physics and Applied Physics, Yale University, New Haven, Connecticut 06520, USA}
\author{K.~B\o rkje}
\affiliation{Departments of Physics and Applied Physics, Yale University, New Haven, Connecticut 06520, USA}
\author{S.~M.~Girvin}
\affiliation{Departments of Physics and Applied Physics, Yale University, New Haven, Connecticut 06520, USA}

\date{\today}

\begin{abstract}
Optomechanics experiments are rapidly approaching the regime where the radiation pressure of a single photon displaces the mechanical oscillator by more than its zero-point uncertainty. We show that in this limit the power spectrum has multiple sidebands and that the cavity response has several resonances in the resolved-sideband limit. Using master-equation simulations, we also study the crossover from the weak-coupling many-photon to the single-photon strong-coupling regime. Finally, we find non-Gaussian steady-states of the mechanical oscillator when multi-photon transitions are resonant. Our study provides the tools to detect and take advantage of this novel regime of optomechanics.
\end{abstract}



\maketitle

\emph{Introduction.} Optomechanics is a rapidly growing field of research studying mechanical degrees of freedom coupled to modes of optical cavities via radiation pressure, optical gradient, or photothermal forces \cite{Kippenberg2008, Marquardt2009}. Work in this area is largely motivated by building more sensitive mass and force sensors \cite{Rugar2004}, providing long-range interaction between qubits in future quantum information hardware \cite{Rabl2010}, and probing quantum mechanics at increasingly large mass and length scales \cite{Marshall2003}.

In the standard optomechanics setup, the position of a mechanical oscillator parametrically modulates the frequency of an optical cavity mode. In most experiments to date this optomechanical coupling is small compared to the mechanical frequency and the cavity linewidth. However, if the cavity is strongly driven and thus contains a large number of photons, the coupling between the mechanical oscillator and the fluctuations of the cavity field is enhanced by a factor $\sqrt{n}$, where $n$ is the mean photon number in the cavity. This has recently led to the observation of radiation-pressure effects, e.g.~red-sideband cooling~\cite{Gigan2006, Schliesser2006, Teufel2008, Thompson2008, Rocheleau2009, Teufel2011}, normal-mode splitting \cite{Groblacher2009, Teufel2010}, and optomechanically-induced transparency \cite{Weis2010, Safavi2010, Teufel2010}.

In this weak coupling regime the Hamiltonian is quadratic so that ordinary thermal and vacuum noise lead to Gaussian steady-states. To create more general and possibly more interesting and useful states one either needs non-Gaussian input noise, e.g.~driving the system with single-photon sources \cite{Akram2010}, or one has to make the system nonlinear. The latter can be achieved either via measurement backaction \cite{Borkje2011} or intrinsic non-linearities, e.g.~coupling the resonator via a qubit to the mechanical oscillator \cite{OConnell2010} or engineering an optomechanical interaction which couples the position squared of the oscillator to the cavity mode \cite{Thompson2008, Sankey2010, Purdy2010, Nunnenkamp2010}.

Several optomechanics setups, using either ultracold atoms in optical resonators \cite{Gupta2007}, optomechanical crystals \cite{Eichenfield2009} or superconducting circuits \cite{Teufel2010}, are approaching the limit where the radiation pressure of a single photon displaces the mechanical oscillator by more than its zero-point uncertainty. In this single-photon strong-coupling regime the full parametric coupling, i.e.~three-wave mixing in the language of quantum optics, has to be taken into account. To date there exists little literature on this subject with the notable exception of Refs.~\cite{Mancini1997, Bose1997, Ludwig2008}.

In this paper we show how to detect this novel regime of optomechanics and exploit the nonlinear spectrum to create non-Gaussian steady-states of the mechanical oscillator. For weak coherent optical drive we use the polaron transformation to calculate properties of the output light to all orders in the optomechanical coupling. We find that the power spectrum has multiple mechanical sidebands and the cavity response has several resonances in the resolved-sideband limit. Using master-equation simulations, we calculate these observables throughout the crossover from the many-photon to the single-photon limit. Finally, we show that multi-photon transitions can lead to non-Gaussian steady-states which might enable the observation of quantum tunneling and noise-induced switching in optomechanical systems.

\begin{figure}
\centering
\includegraphics[width=\columnwidth]{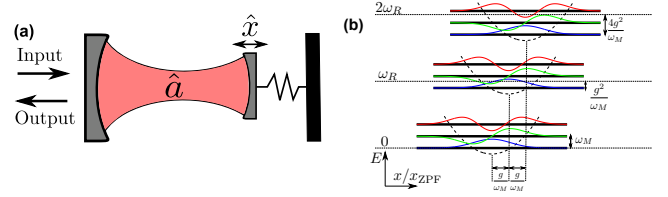}
\caption{(Color online) (a) Standard optomechanics setup: the position $\hat{x}$ of a mechanical oscillator is parametrically coupled to a driven cavity mode $\hat{a}$. (b) Spectrum and eigenfunctions of Hamiltonian (\ref{Ham1}). The energy axis is not to scale. Parabolas indicate the displaced harmonic oscillator potentials for $n=0,1,2$ photons and $g<0$.}
\label{fig:eigensystem}
\end{figure}

\emph{Model.} We consider the standard model of optomechanical systems where the position of a mechanical oscillator, $\hat{x} = x_\mathrm{ZPF}(\des{b} + \cre{b})$, is parametrically coupled to an optical cavity mode $\des{a}$, see Fig.~\ref{fig:eigensystem} (a). Setting $\hbar = 1$ the Hamiltonian reads
\begin{equation}
\hat{H}_0 = \omega_R \cre{a} \des{a} + \omega_M \cre{b} \des{b} + g \cre{a} \des{a} (\des{b} + \cre{b})
\label{Ham1}
\end{equation}
where $\omega_R$ is the resonator frequency, $\omega_M$ the mechanical frequency, and $g=\omega_R' x_\mathrm{ZPF}$ is the optomechanical coupling. $x_\mathrm{ZPF} = (2M\omega_M)^{-1/2}$ is the zero-point uncertainty, $M$ the mass of the mechanical oscillator, and $\omega_R' = \frac{\partial \omega_R}{\partial x}$ the derivative of the resonator energy with respect to the oscillator position $x$. $\des{a}$ and $\des{b}$ are bosonic annihilation operators for the cavity mode and the mechanical oscillator, respectively.

Note first that the Hamiltonian (\ref{Ham1}) conserves photon number, i.e.~$[ \cre{a} \des{a}, \hat{H}_0] = 0$. The Hamiltonian in the subspace of $n$ photons is a harmonic oscillator with frequency $\omega_M$ which is displaced by $-n x_0$, where $x_0 = -2 x_\mathrm{ZPF} g/\omega_M$ is the displacement caused by one photon. Thus, the eigenvalues of (\ref{Ham1}) are $E_{nm} = \omega_R n - g^2n^2/\omega_M + \omega_M m$ with non-negative integers $n$ and $m$. The anharmonicity is proportional to the product of photon number $n$ and oscillator displacement which is linear in the photon number $n$. We show the spectrum and eigenfunctions of Hamiltonian (\ref{Ham1}) in Fig.~\ref{fig:eigensystem} (b).

In order to include drive and decay we use standard input-output theory \cite{GirvinRMP}. In a frame rotating at the frequency of the optical drive, the non-linear quantum Langevin equations read
\begin{align}
\dot{\des{a}} &= + i \Delta \des{a} - \frac{\kappa}{2} \des{a} - i g \left(\cre{b} + \des{b} \right) \des{a} + \sqrt{\kappa} \, \des{a}_\mathrm{in} \label{motion1} \\
\dot{\des{b}} &= -i \omega_M \des{b} - \frac{\gamma}{2} \des{b} - i g \cre{a} \des{a} + \sqrt{\gamma} \, \des{b}_\mathrm{in}. \label{motion2}
\end{align}
where $\Delta = \omega_L - \omega_R$ is the detuning between laser $\omega_L$ and resonator frequency $\omega_R$, and $\gamma$ and $\kappa$ are the mechanical and cavity damping rates. The cavity input $\des{a}_\mathrm{in}$ is a sum of a coherent amplitude $\bar{a}_\mathrm{in}$ and a vacuum noise operator $\des{\xi}$ satisfying $\langle \des{\xi}(t) \cre{\xi}(t') \rangle = \delta(t-t')$ and $\langle \cre{\xi}(t) \des{\xi}(t') \rangle = 0$. Finally, we assume that the mechanical bath is Markovian and has a temperature $T$, i.e.~$\langle \des{b}_\mathrm{in}(t) \cre{b}_\mathrm{in}(t') \rangle = (n_\mathrm{th}+1) \delta(t-t')$ and $\langle \cre{b}_\mathrm{in}(t) \des{b}_\mathrm{in}(t') \rangle = n_\mathrm{th} \delta(t-t')$ with $n_\mathrm{th}^{-1} = e^{\hbar \omega_M/k_B T}-1$.
 
The model is characterized by three dimensionless parameters: the mechanical quality factor $\omega_M/\gamma$, the resolved-sideband parameter $\omega_M / \kappa$, and the granularity parameter $g/\kappa$ \cite{Ludwig2008, Murch2008}. The latter is the cavity frequency shift in units of its linewidth when the oscillator is displaced by one zero-point uncertainty $x_\mathrm{ZPF}$. Finally, $2g/\omega_M = 2(g/\kappa) \times (\omega_M / \kappa)^{-1}$ is the oscillator displacement in units of $x_\mathrm{ZPF}$ caused by the radiation pressure of a single photon. If $|g|\ge \omega_M$ we will say the system is in the single-photon strong-coupling regime.

\begin{figure}
\centering
\includegraphics[width=0.45\columnwidth]{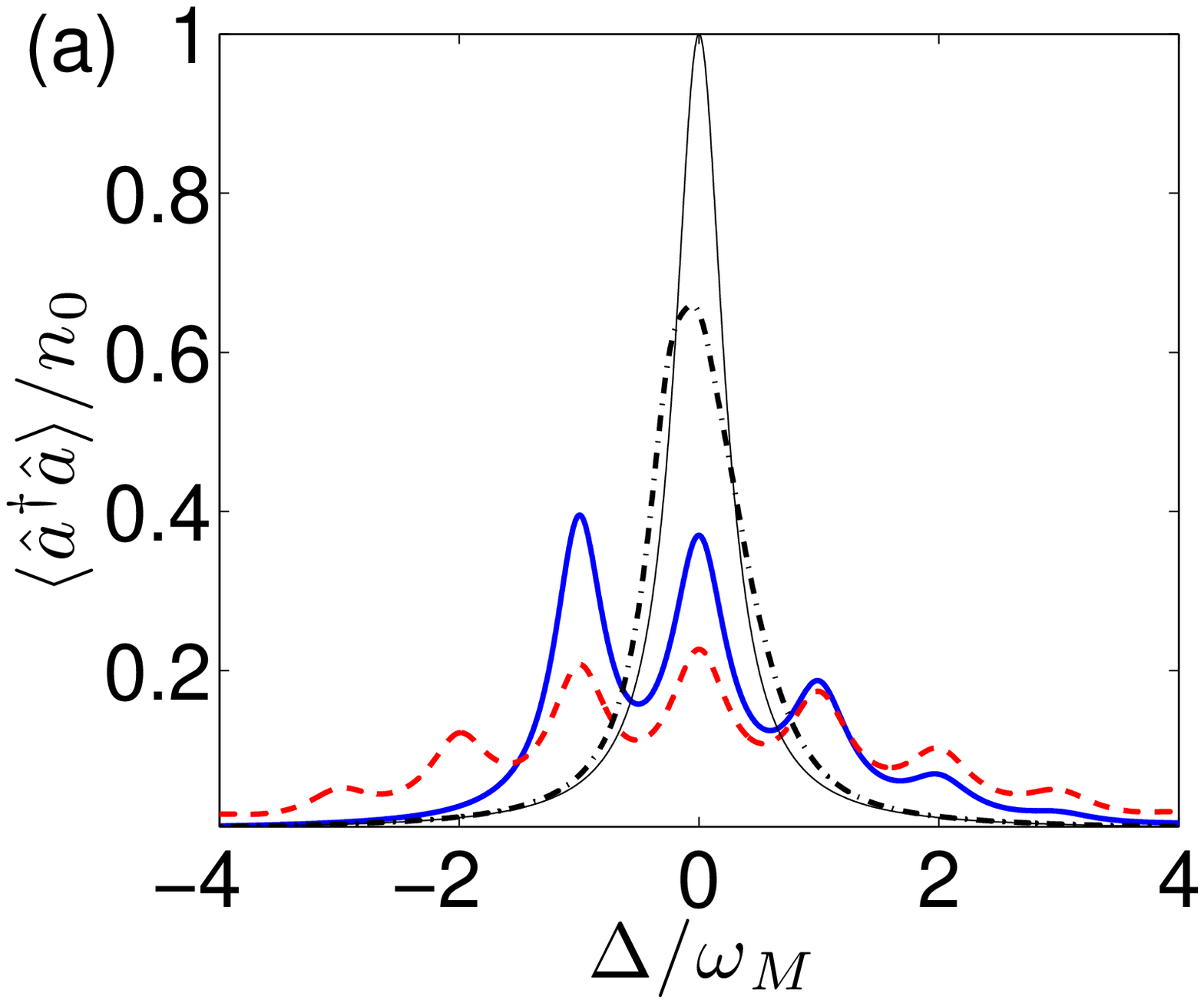}
\includegraphics[width=0.45\columnwidth]{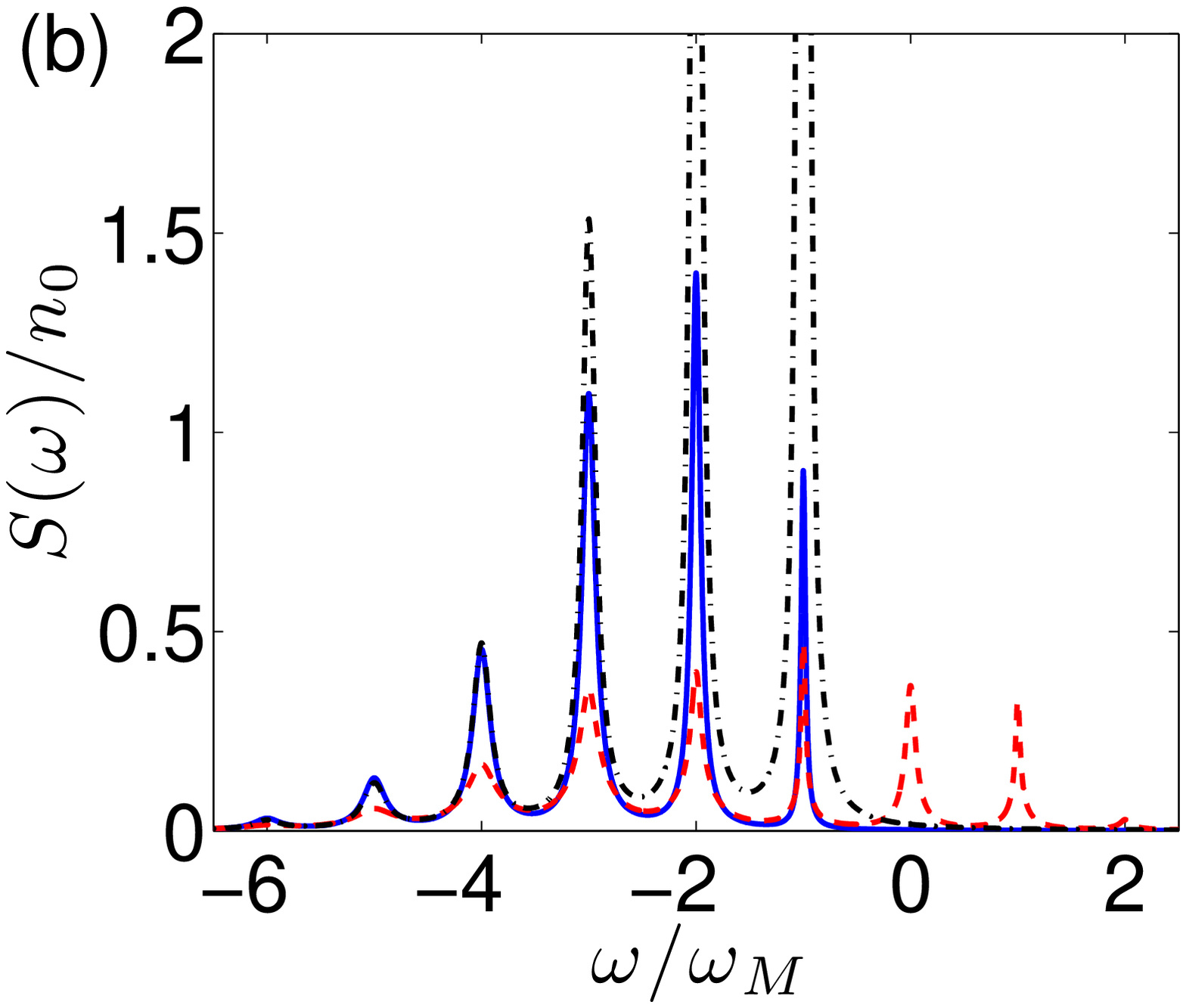}
\caption{(Color online) Detecting the single-photon strong-coupling regime: (a) Steady-state mean photon number $\langle \cre{a} \des{a} \rangle$ as a function of detuning $\Delta$ and (b) power spectrum $S(\omega)$ at $\Delta = 0$. $\omega_M = g$ and $\omega_M/\gamma = 20$ for all curves, $\omega_M/\kappa = 2$ and $n_\mathrm{th}=0$ (blue solid), $\omega_M/\kappa = 2$ and $n_\mathrm{th}=1$ (red dashed) as well as $\omega_M/\kappa = 0.5$ and $n_\mathrm{th}=0$ (black dash-dotted). The thin black solid line in (a) shows the empty cavity response $g=0$ for comparison. $n_0$ is the mean photon number on resonance, i.e.~$ n_0 = 4 \Omega^2 / \kappa^2$.}
\label{fig:spectra}
\end{figure}

\emph{Approximate solution for weak drive.} It is well known that the Hamiltonian $\hat{H}_0$ can be diagonalized by the polaron transform  given by $\hat{U} = e^{-\hat{S}}$ with $\hat{S} = \frac{g}{\omega_M} \cre{a} \des{a} (\cre{b} - \des{b})$ \cite{Mahan2000}.
Here we use it to find an approximate solution to Eqs.~(\ref{motion1}) and (\ref{motion2}).

In steady-state and for weak optical drive we obtain
\begin{equation}
\hat{a}(t) = \sqrt{\kappa} \int_{-\infty}^t \!\!\!\!\! d\tau \, e^{-(\kappa/2-i\tilde{\Delta})(t-\tau)} e^{\hat{X}(t)} e^{-\hat{X}(\tau)} \des{a}_\mathrm{in}(\tau)
\end{equation}
where we defined $\tilde{\Delta} = \Delta + g^2/\omega_M$ and $\hat{X}(t)$ is given by 
\begin{equation}
\hat{X}(t) = \frac{\sqrt{\gamma} g}{\omega_M}  \int_{-\infty}^t \!\!\!\!\! d \tau \, e^{-\gamma(t-\tau)/2} \left[ e^{-i \omega_M (t-\tau)} \des{b}_\mathrm{in}(\tau) - \mathrm{H.c.} \right]
\end{equation}

Using this analytic approach, we calculate properties of the optical field. We get for the steady-state mean photon number
\begin{align}
\frac{\langle \cre{a}\des{a} \rangle}{n_0} = &\sum_{n = 0}^\infty \frac{(g/\omega_M)^{2n}}{4n!} \sum_{k = 0}^n {n \choose k} (n_\mathrm{th}+1)^{n-k} n_\mathrm{th}^k \nonumber \\
&\times \frac{\kappa(\kappa+n\gamma)e^{-(g/\omega_M)^2(2n_\mathrm{th}+1)}}{(\frac{\kappa+n\gamma}{2})^2 + (\tilde{\Delta} - (n-2k)\omega_M)^2}
\label{number}
\end{align}
and the cavity amplitude relevant for homodyne experiments
\begin{align}
\frac{\langle \des{a} \rangle}{\sqrt{n_0}} = &\sum_{n = 0}^\infty \frac{(g/\omega_M)^{2n}}{2n!} \sum_{k = 0}^n {n \choose k} (n_\mathrm{th}+1)^{n-k} n_\mathrm{th}^k \nonumber \\
&\times \frac{\kappa e^{-(g/\omega_M)^2(2n_\mathrm{th}+1)}}{(\frac{\kappa+n\gamma}{2}) - i (\tilde{\Delta} - (n-2k)\omega_M)}
\label{homodyne}
\end{align}
where $n_0 = 4\Omega^2/\kappa^2$ with $\Omega = \sqrt{\kappa} |\bar{a}_\mathrm{in}|$ is the mean photon number for $g=0$ on resonance $\Delta = 0$. The quantities (\ref{number}) and (\ref{homodyne}) are sums of resonances which are spaced by the mechanical frequency $\omega_M$. Let us discuss first the case of zero temperature, i.e.~$n_\mathrm{th}=0$, when only terms with $k = 0$ contribute in Eqs.~(\ref{number}) and (\ref{homodyne}). In this case, the resonances are weighted by a Poisson distribution with variance $(g/\omega_M)^2$ and the widths are $\kappa + n\gamma$. The resonances can be understood as transitions between the vacuum state $\ket{0,0}$ and the manifold of one-photon eigenstates $\ket{1,m}$ of the Hamiltonian (\ref{Ham1}). They are resonant if the laser frequency $\omega_L$ matches $\omega_L = E_{1m} - E_{00} = \omega_R -g^2/\omega_M + m \omega_M$. The Poission distribution is due to the Franck-Condon factors $|\bra{m} e^{\hat{X}} \ket{0}|^2 = |\int dx \, \varphi_m^*(x-x_0) \varphi_0(x)|^2 =(g/\omega_M)^{2m} e^{-(g/\omega_M)^2}/m!$ where $\ket{m}$ is the state with $m$ phonons, $\varphi_m(x)$ is its real-space wavefunction, and $x_0/x_\mathrm{ZPF} = 2g/\omega_M$. At finite temperature the states $\ket{0,m}$ with $m>0$ are thermally occupied leading to a redistribution of weight among the peaks and additional resonances at $\omega_L = \omega_R -g^2/\omega_M - m \omega_M$.

In the limit $\kappa \gg \gamma$ we obtain the cavity spectrum $S(\omega) = \int_{-\infty}^\infty dt \, e^{i\omega t} [\langle \cre{a}(t) \des{a}(0)\rangle - |\langle \des{a}(t)\rangle|^2]$ as
\begin{align}
S (\omega) = \sum_{m,n = 0}^\infty \frac{C_{mn} (m+n) \gamma}{\left[\frac{(m+n)\gamma}{2}\right]^2 + \left[(\omega - (m-n)\omega_M\right]^2}
\label{spectrum}
\end{align}
with the $n=m=0$ term excluded. The coefficients $C_{mn}$ are independent of $\omega$ but rather involved and will not be shown.

The optical output spectrum has sideband peaks at integer multiples of the mechanical frequency $\omega_M$ whose widths are multiples of the mechanical linewidth $\gamma$. At zero temperature there are peaks only at negative frequencies because photons can only create phonons and leave the cavity with frequencies smaller than the laser frequency $\omega_L$. At finite temperature (or stronger optical drive) additional peaks appear at positive frequencies since there is a finite probability that a photon absorbs the energy of one or more phonons and leaves the cavity with a frequency larger than the laser frequency $\omega_L$. In passing we note that driving on these additional sidebands in the resolved-sideband limit leads to multi-phonon cooling which will be discussed in a future publication.

In Fig.~\ref{fig:spectra} (a) we plot the steady-state mean photon number $\langle \cre{a} \des{a} \rangle$ as a function of detuning $\Delta$ for a system entering the single-photon strong-coupling regime, $g = \omega_M$. In the good-cavity limit $\kappa < \omega_M$ the cavity response shows several resolved resonances. At finite thermal phonon number additional peaks appear and their weights are redistributed until eventually they blur into a broad thermal background. In the bad-cavity limit $\kappa > \omega_M$ the resonances overlap and broaden the empty cavity resonance. In Fig.~\ref{fig:spectra} (b) we present the output spectrum $S(\omega)$ at zero detuning $\Delta = 0$. It shows a series of peaks at multiples of the mechanical frequency $\omega_M$ for all sets of parameters: within and outside the resolved-sideband limit as well as for a zero and finite thermal phonon number.

\begin{figure}
\centering
\includegraphics[width=0.3\columnwidth]{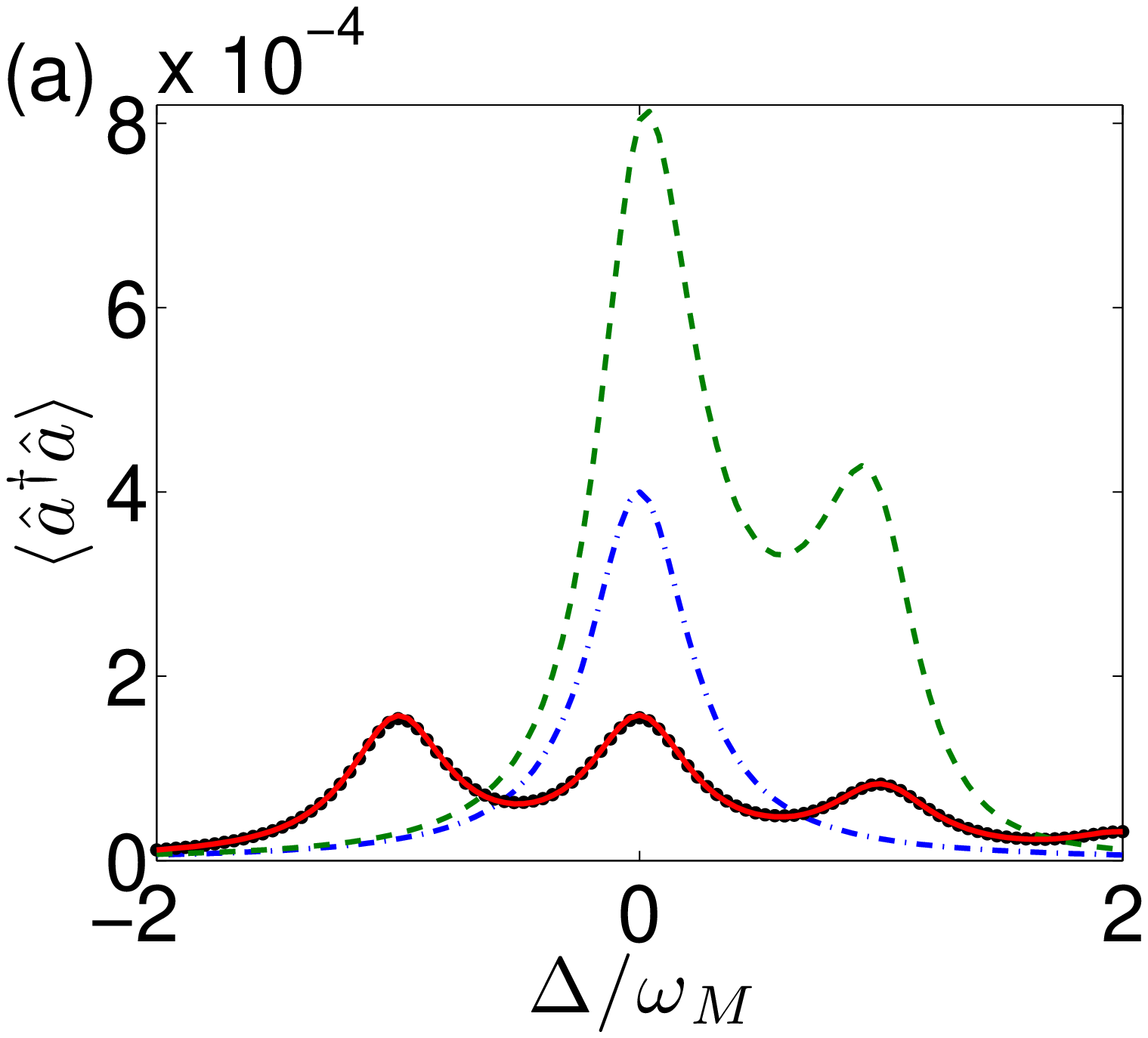}
\includegraphics[width=0.3\columnwidth]{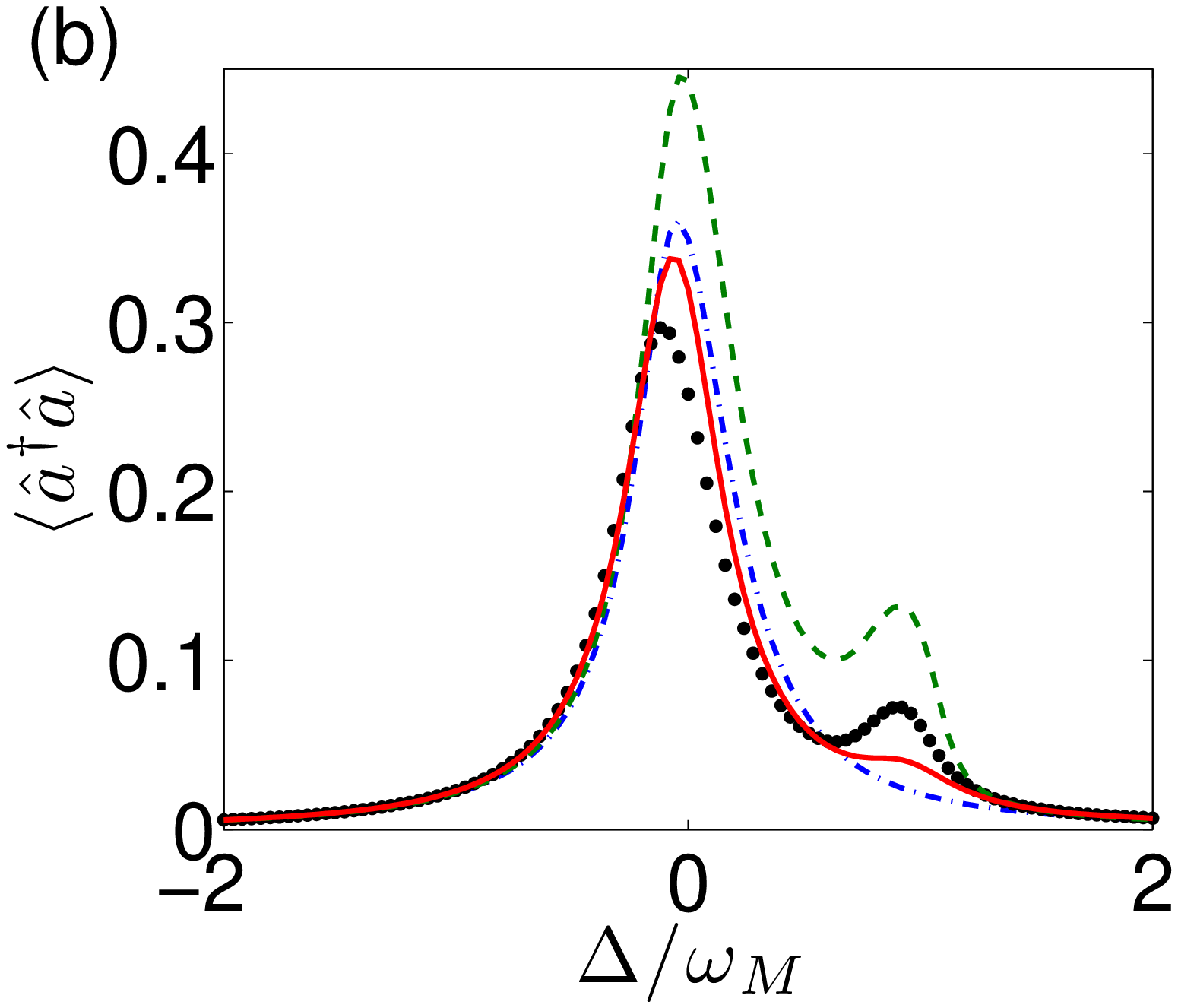}
\includegraphics[width=0.3\columnwidth]{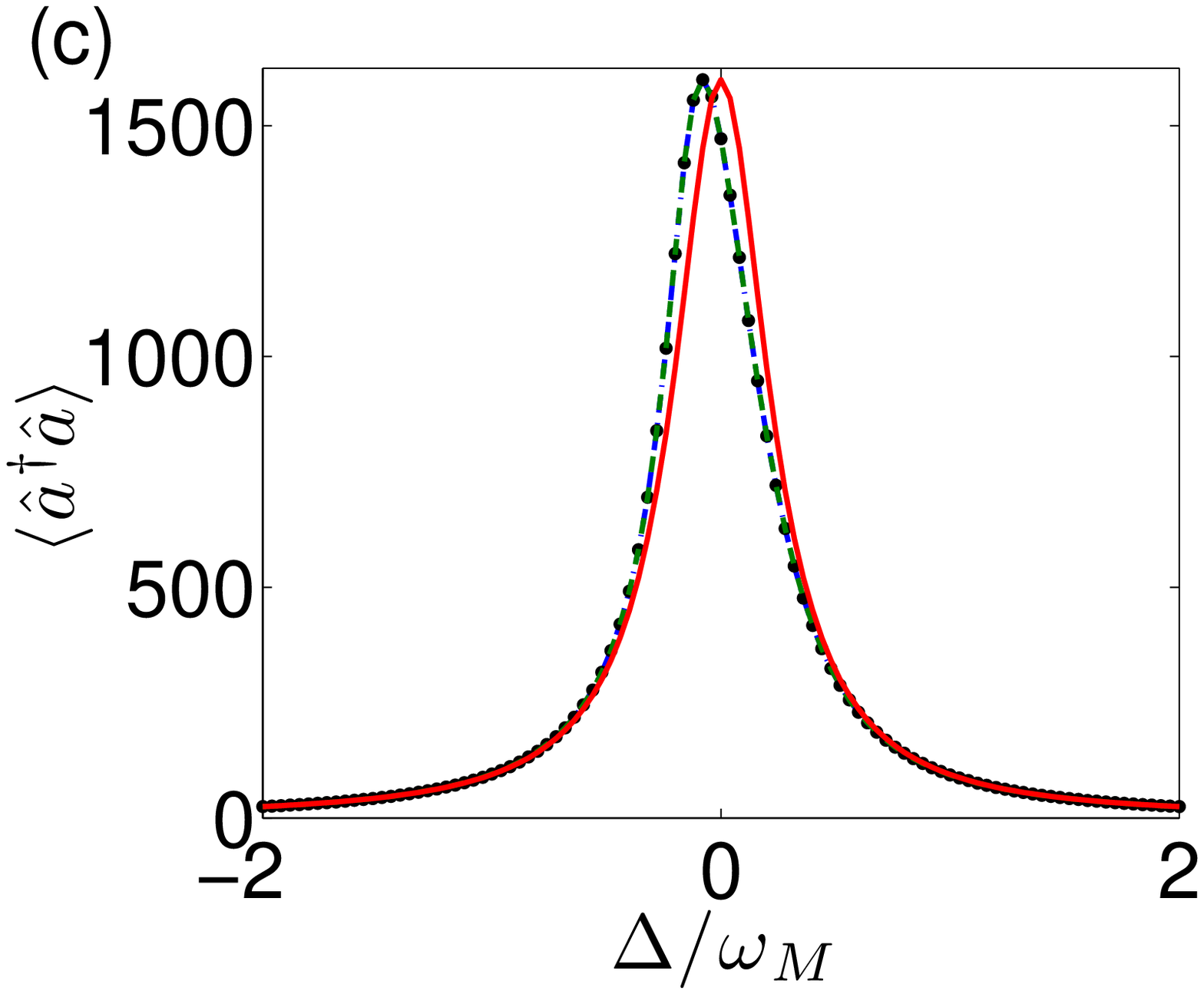}
\includegraphics[width=0.3\columnwidth]{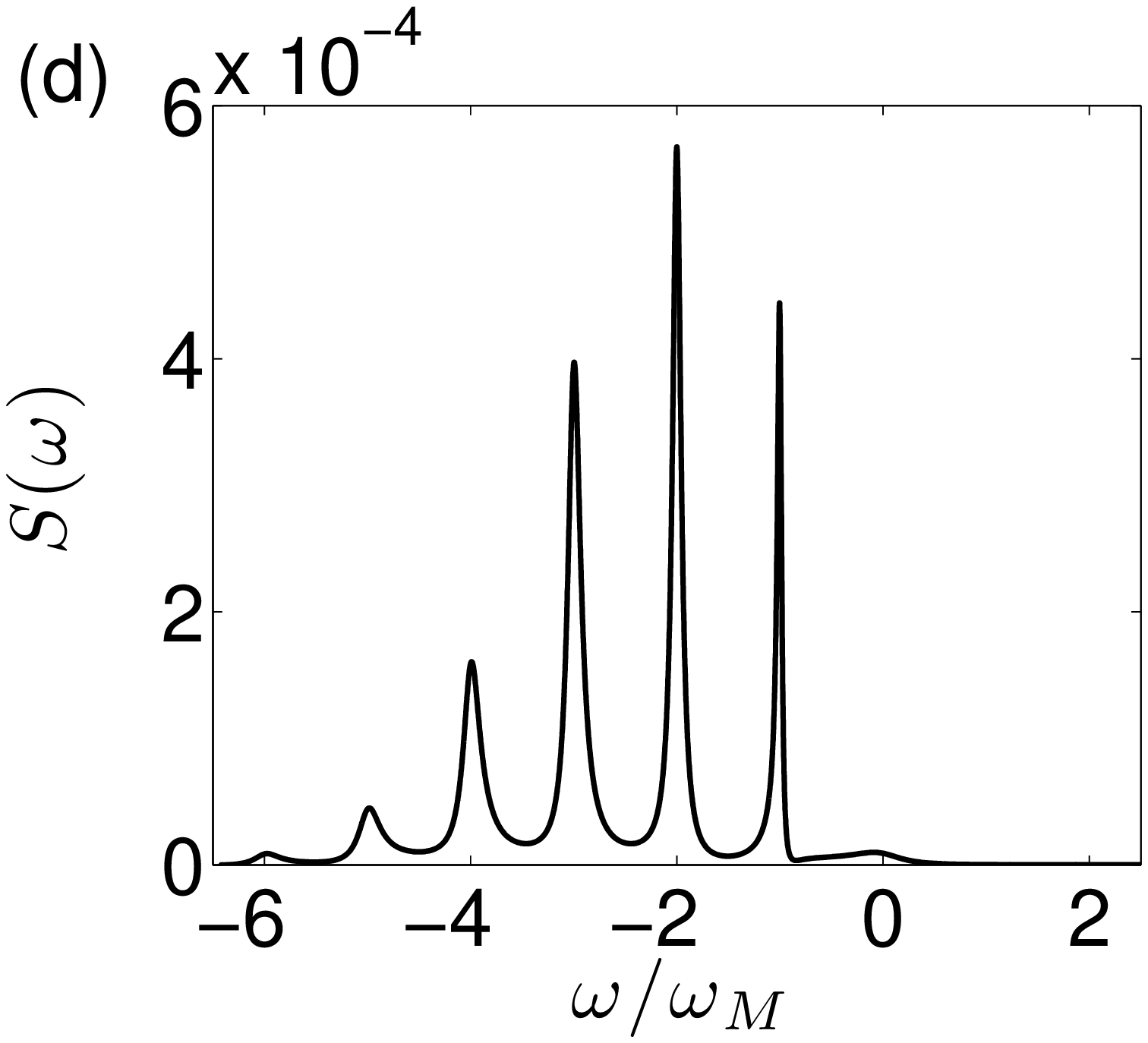}
\includegraphics[width=0.3\columnwidth]{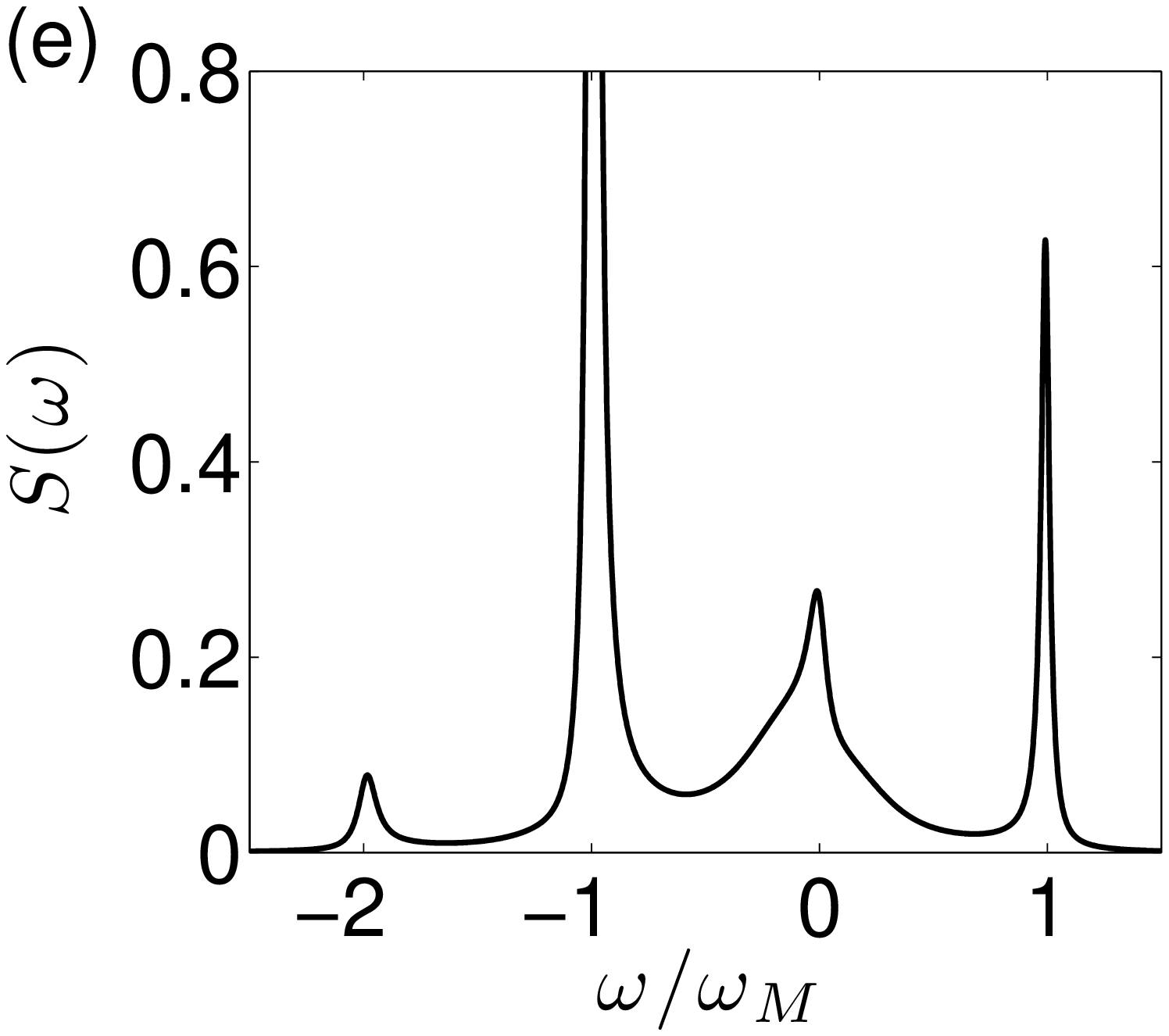}
\includegraphics[width=0.3\columnwidth]{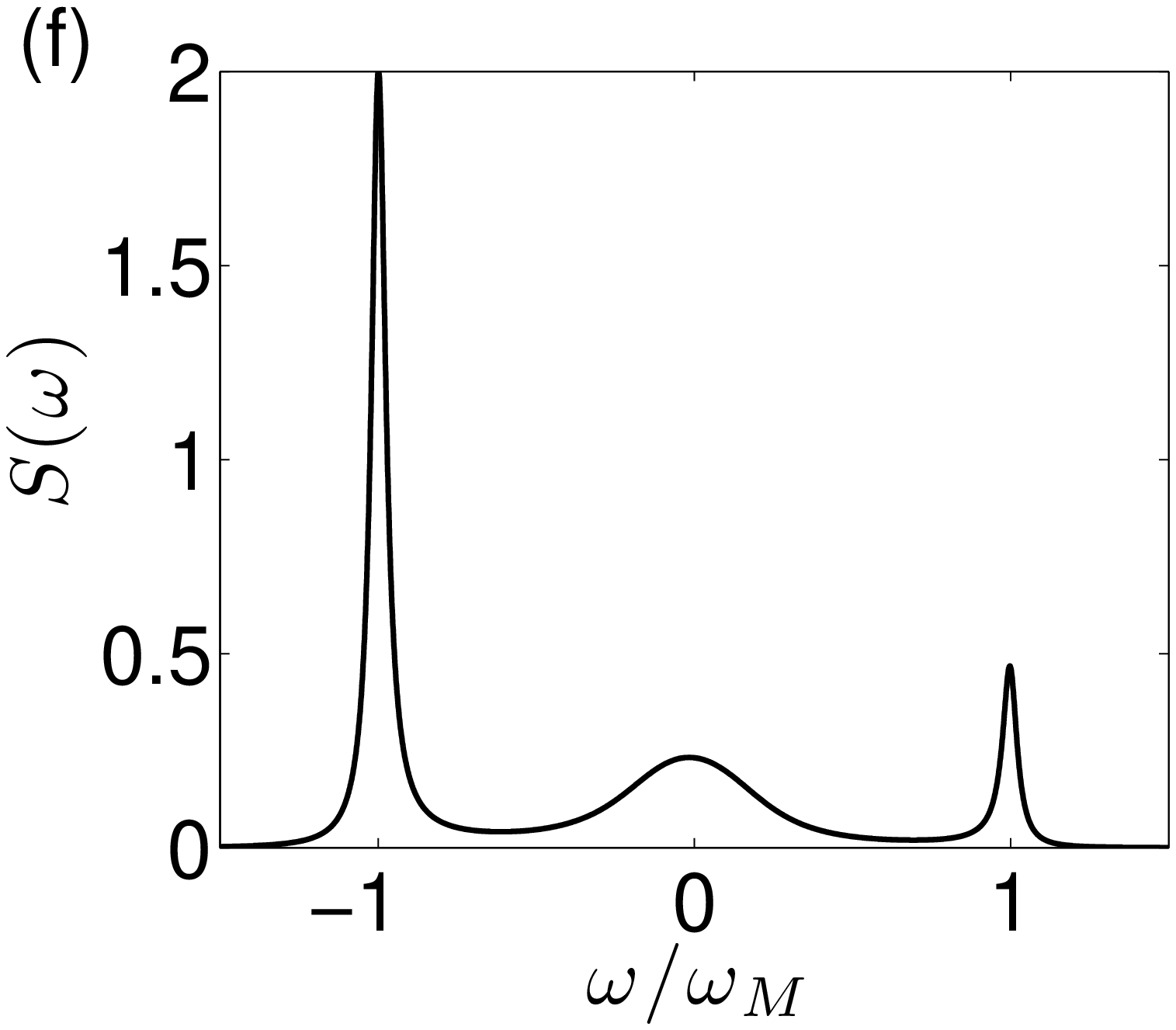}
\caption{(Color online) Crossover from many- to single photon limit: Steady-state mean photon number $\langle \cre{a} \des{a} \rangle$ as a function of detuning $\Delta$. (a) $\Omega/\kappa = 0.01$ and $g/\kappa = 2$, (b) $\Omega/\kappa = 0.5$ and $g/\kappa = 0.5$, and (c) $\Omega/\kappa = 20$ and $g/\kappa = 0.01$. Parameters are $\omega_M/\kappa = 2$, $\Delta/\kappa = -2$, $\omega_M/\gamma=100$ and $n_\mathrm{th}=0$. We show Eq.~(\ref{number}) (red solid), $|\bar{a}|^2$ (blue dash-dotted), linear theory $|\bar{a}|^2 + \langle \cre{d} \des{d} \rangle$ (green dashed) and simulations of Eq.~(\ref{master2}) (black dots). (d-f) Output spectrum $S(\omega)$ at $\Delta = 0$ for the same parameters from simulations of Eq.~(\ref{master2}).}
\label{fig:crossover}
\end{figure}

\emph{Crossover between the many- and the single-photon limit.} Let us now compare the single-photon strong-coupling regime to the more familiar case of weak optomechanical coupling and study the crossover between these two extreme limits.

For numerical simulations it is advantageous to use the displacement transformation by writing $\des{a} = \bar{a} + \des{d}$ and $\des{b} = \bar{b} + \des{c}$. We obtain a set of coupled equations for the mean values $\bar{a}$ and $\bar{b}$: $0 = i \Delta \bar{a} - \frac{\kappa}{2} \bar{a} - i \Omega - i g (\bar{b} + \bar{b}^*) \bar{a}$ and $0 = -i \omega_M \bar{b} - \frac{\gamma}{2} \bar{b} - i g |\bar{a}|^2$. It is well-known that these nonlinear equations have either one or three solutions. In the latter case the system is said to be (classically) bistable. The operators $\des{d}$ and $\des{c}$ describing the fluctuations around the mean values $\bar{a}$ and $\bar{b}$, respectively, satisfy equations of motion equivalent to the quantum master equation
\begin{equation}
\dot{\varrho} = -i \left[ \hat{H}'_0, \varrho \right] + \kappa \mathcal{D}[\des{d}] \varrho + \gamma(n_\mathrm{th}+1) \mathcal{D}[\des{c}] \varrho + \gamma n_\mathrm{th} \mathcal{D}[\cre{c}] \varrho
\label{master2}
\end{equation}
with the Hamiltonian
\begin{equation}
\hat{H}'_0 = - \Delta' \cre{d} \des{d} + \omega_M \cre{c} \des{c} + g (\bar{a}^* \des{d} + \bar{a} \cre{d}) \left( \des{c} + \cre{c} \right) + g \cre{d} \des{d} \left( \des{c} + \cre{c} \right)
\label{Ham2}
\end{equation}
where the detuning is renormalized $\Delta' = \Delta - g (\bar{b} + \bar{b}^*)$. $\mathcal{D}[\des{o}] \varrho = \des{o} \varrho \cre{o} - (\cre{o} \des{o} \varrho + \varrho \cre{o} \des{o})/2$ is the standard dissipator in Lindblad form. This is an exact description of the system in a frame where the mean of both harmonic oscillators has been displaced to the vacuum.

Outside the bistable region, for large mean cavity amplitude $\bar{a}$ and small optomechanical coupling $g$, the last term in the Hamiltonian (\ref{Ham2}) can be neglected. We then obtain a quadratic Hamiltonian or equivalently a set of linear quantum Langevin equations which can be solved exactly. In this linear theory we have $\langle \des{d} \rangle = 0$, and the photon number is given by $|\bar{a}|^2 +\langle \cre{d} \des{d} \rangle$.

We now compare the predictions of the numerical solution of the quantum master equation (\ref{master2}) to the linear theory and the analytic expressions (\ref{number}) and (\ref{spectrum}) derived above. In Fig.~\ref{fig:crossover} (a-c) we plot the steady-state mean photon number $\langle \cre{a} \des{a} \rangle$ as a function of detuning $\Delta$ for three different sets of parameters. For $\Omega/\kappa = 0.01$ and $g/\kappa = 2$ we are in the single-photon strong-coupling limit. The numerical solution of Eq.~(\ref{master2}) shows several resonances and agrees very well with the analytical expression (\ref{number}). The linear theory is not appropriate in this regime. This is signaled by the fact that the size of the fluctuations by far exceeds the mean photon number: $|\bar{a}|^2 \ll \langle \cre{d} \des{d} \rangle$. At intermediate coupling and drive, $\Omega/\kappa = 0.5$ and $g/\kappa = 0.5$, the numerical simulation of (\ref{master2}) predicts one large peak slightly below $\Delta = 0$ and a small resonance close to the blue sideband $\Delta = \omega_M$. Eq.~(\ref{number}) and the linear theory qualitatively describe this feature but fail to agree with the numerics quantitatively. Finally, for $\Omega/\kappa = 20$ and $g/\kappa = 0.01$ we are well inside the regime where the linear theory is valid. It correctly predicts a slightly asymmetric peak close to $\Delta = 0$.

In Fig.~\ref{fig:crossover} (d-f) we show the optical output spectra $S(\omega)$ for the same parameters obtained from simulations of Eq.~(\ref{master2}). In the single-photon strong-coupling limit it has multiple sidebands and agrees quantitatively with Eq.~(\ref{spectrum}). As the drive strength increases, additional sidebands at positive frequencies appear. With decreasing optomechanical coupling $g$ the weight gradually concentrates in the two sidebands at $\omega = \pm \omega_M$ as predicted by the linear theory.

\begin{figure}
\centering
\includegraphics[width=0.45\columnwidth]{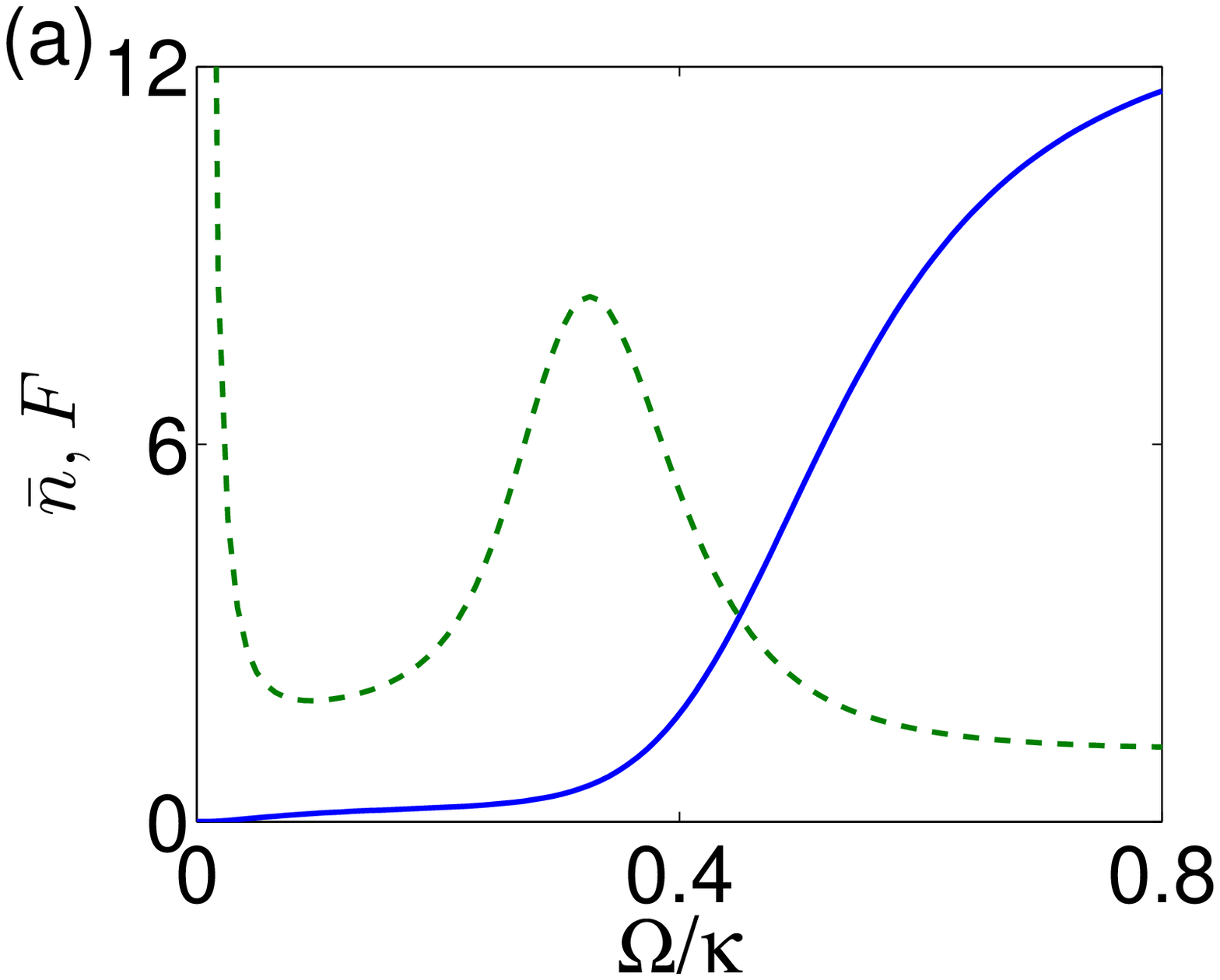}
\includegraphics[width=0.45\columnwidth]{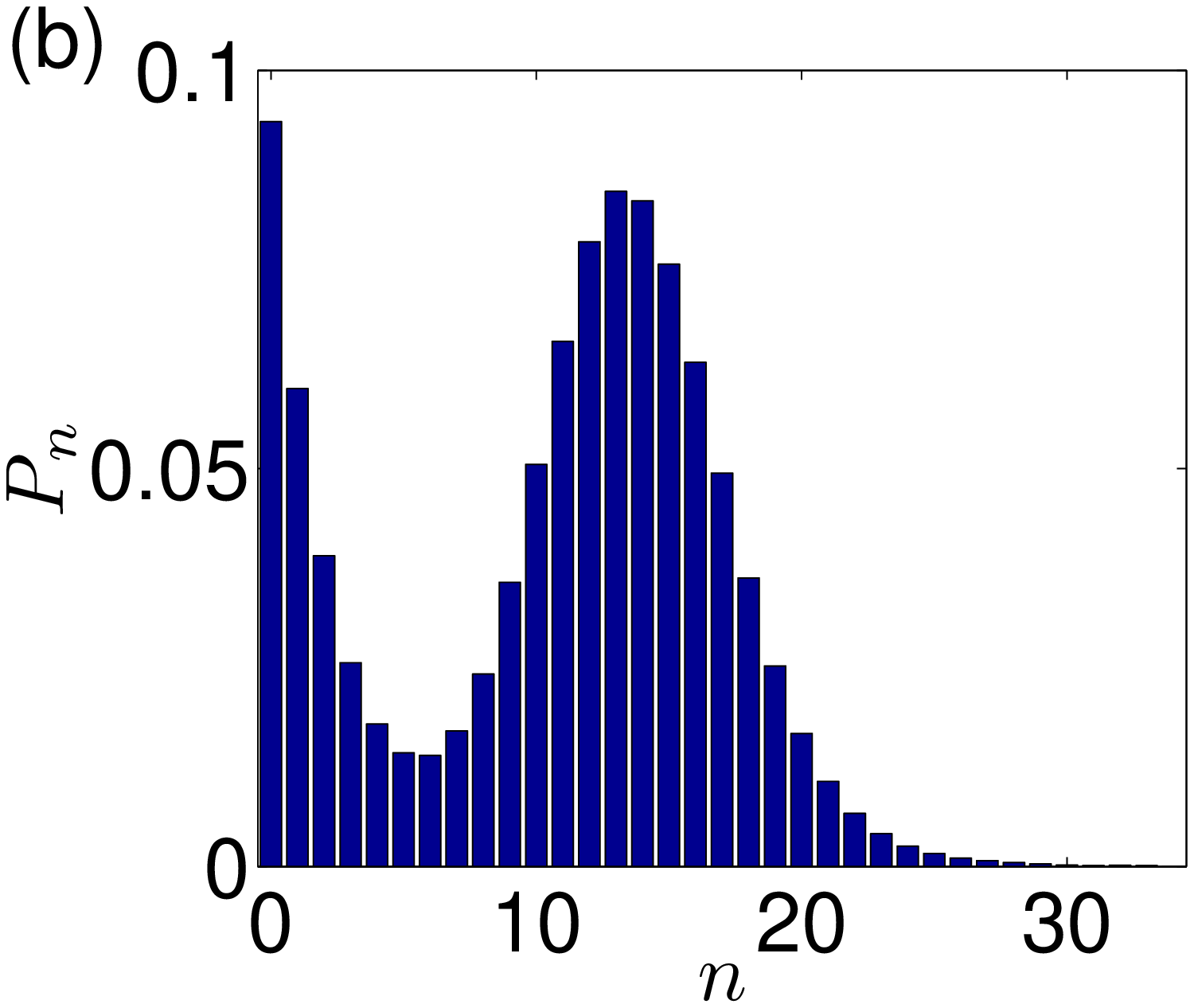}
\caption{(Color online) Non-Gaussian steady-states via multi-photon transitions. (a) Steady-state mean phonon number $\langle\cre{b} \des{b} \rangle$ (blue solid) and the second-order coherence of the mechanical oscillator $F=\langle \cre{b} \cre{b} \des{b} \des{b} \rangle / (\langle \cre{b} \des{b} \rangle)^2$ (green dashed) as a function of drive strength $\Omega$. (b) Phonon number distribution $P_n$ at $\Omega/\kappa = 0.6$. Parameters are $\Delta = -3g^2/\omega_M$, $\omega_M/\kappa = 2$, $\omega_M/\gamma = 1000$, and $g/\kappa = 1$.}
\label{fig:bistable}
\end{figure}

\emph{Non-Gaussian steady-states via multi-photon transitions.} In the final part of this paper we address the question as to how the nonlinear single-photon strong-coupling limit leads to non-Gaussian steady-states for the mechanical oscillator.

Recalling the spectrum of the Hamiltonian (\ref{Ham1}) we notice that for $\Delta = -ng^2/\omega_M$ multi-photon transitions between the vacuum state $\ket{0,0}$ and the lowest-energy state with $n$ photons $\ket{n,0}$ are resonant, i.e.~$E_{00} + n \omega_L = E_{n0}$. As all intermediate transitions are off-resonant, the system is in the Franck-Condon blockade regime \cite{Koch2005}, and we expect that at weak drive the system will stay close to the vacuum state and a strong drive induces multi-photon transitions.

In Fig.~\ref{fig:bistable} (a) we show the steady-state mean phonon number $\bar{n} = \langle\cre{b} \des{b} \rangle$ and the second-order coherence of the mechanical oscillator $F=\langle \cre{b} \cre{b} \des{b} \des{b} \rangle / (\langle \cre{b} \des{b} \rangle)^2$ as a function of drive strength $\Omega$. For small drive $\Omega/\kappa < 0.3$ the mean phonon number $\bar{n}$ remains small as expected. After a region of strong number fluctuations $F \gg 1$, the phonon number increases rapidly with drive strength $\Omega$. We note that the phonon number fluctuations also become large in the limit of small drive $\Omega \rightarrow 0$ which has been discussed in Ref.~\cite{Koch2005}.

Since $|\langle \des{b}\rangle|$ is small for all drive strengths $\Omega$ considered here (not shown), the Wigner function of the mechanical oscillator is rotationally invariant and the phonon number distribution $P_n$ contains the complete information of the reduced density matrix. In Fig.~\ref{fig:bistable} (b) we plot the phonon number distribution $P_n$ for $\Omega/\kappa = 0.6$. We can clearly distinguish two peaks: one at zero and one at $n=14$ phonons. With increasing drive strength $\Omega$ weight is gradually transferred from the former to the latter. We interpret this as a statistical mixture of two different oscillation amplitudes and expect the system to exhibit quantum tunneling and noise-induced switching \cite{Rigo1997}. We note that similar states have been reported in Ref.~\cite{Ludwig2008}.

\emph{Conclusion.} Motivated by recent experiments, we explored optomechanics in the regime where the radiation pressure of a single photon displaces the mechanical oscillator by more than its zero-point uncertainty. We demonstrated that the output spectrum and cavity response are qualitatively modified and showed how to create non-Gaussian steady states of the mechanical oscillator. Our study opens many further questions about the single-photon strong-coupling regime whose physics is far from well-understood, including e.g.~the fate of red-sideband cooling and ponderomotive squeezing.

\emph{Acknowledgements.} We would like to thank Jens Eisert and Dan M.~Stamper-Kurn for insightful discussions and acknowledge support from NSF under Grant No.~DMR-1004406 (AN and SMG) and DMR-0653377 (SMG) as well as from the Research Council of Norway under Grant No.~191576/V30 (KB). Part of the calculations were performed with the Quantum Optics Toolbox \cite{Tan1999}.

\emph{Note added.} During the final stages of this project, a related paper by Rabl appeared \cite{Rabl2011}.


\begin{thebibliography}{30}
\expandafter\ifx\csname natexlab\endcsname\relax\def\natexlab#1{#1}\fi
\expandafter\ifx\csname bibnamefont\endcsname\relax
  \def\bibnamefont#1{#1}\fi
\expandafter\ifx\csname bibfnamefont\endcsname\relax
  \def\bibfnamefont#1{#1}\fi
\expandafter\ifx\csname citenamefont\endcsname\relax
  \def\citenamefont#1{#1}\fi
\providecommand{\bibinfo}[2]{#2}
\providecommand{\eprint}[2][]{\url{#2}}

\bibitem[{\citenamefont{Kippenberg and Vahala}(2008)}]{Kippenberg2008}
\bibinfo{author}{\bibfnamefont{T.~J.} \bibnamefont{Kippenberg}}
  \bibnamefont{and} \bibinfo{author}{\bibfnamefont{K.~J.}
  \bibnamefont{Vahala}}, \bibinfo{journal}{Science}
  \textbf{\bibinfo{volume}{321}}, \bibinfo{pages}{1172} (\bibinfo{year}{2008}).

\bibitem[{\citenamefont{Marquardt and Girvin}(2009)}]{Marquardt2009}
\bibinfo{author}{\bibfnamefont{F.}~\bibnamefont{Marquardt}} \bibnamefont{and}
  \bibinfo{author}{\bibfnamefont{S.~M.} \bibnamefont{Girvin}},
  \bibinfo{journal}{Physics} \textbf{\bibinfo{volume}{2}}, \bibinfo{eid}{40}
  (\bibinfo{year}{2009}).

\bibitem[{\citenamefont{Rugar et~al.}(2004)\citenamefont{Rugar, Budakian,
  Mamin, and Chui}}]{Rugar2004}
\bibinfo{author}{\bibfnamefont{D.}~\bibnamefont{Rugar}},
  \bibinfo{author}{\bibfnamefont{R.}~\bibnamefont{Budakian}},
  \bibinfo{author}{\bibfnamefont{H.~J.} \bibnamefont{Mamin}}, \bibnamefont{and}
  \bibinfo{author}{\bibfnamefont{B.~W.} \bibnamefont{Chui}},
  \bibinfo{journal}{Nature (London)} \textbf{\bibinfo{volume}{430}}, \bibinfo{pages}{329} (\bibinfo{year}{2004}).

\bibitem[{\citenamefont{Rabl et~al.}(2010)\citenamefont{Rabl, Kolkowitz,
  Koppens, Harris, Zoller, and Lukin}}]{Rabl2010}
\bibinfo{author}{\bibfnamefont{P.}~\bibnamefont{Rabl}} \textit{et~al.},
  \bibinfo{journal}{Nature Physics} \textbf{\bibinfo{volume}{6}},
  \bibinfo{pages}{602} (\bibinfo{year}{2010}).

\bibitem[{\citenamefont{Marshall et~al.}(2003)\citenamefont{Marshall, Simon, Penrose, and Bouwmeester}}]{Marshall2003}
\bibinfo{author}{\bibfnamefont{W.}~\bibnamefont{Marshall}},
  \bibinfo{author}{\bibfnamefont{C.}~\bibnamefont{Simon}},
  \bibinfo{author}{\bibfnamefont{R.}~\bibnamefont{Penrose}}, \bibnamefont{and}
  \bibinfo{author}{\bibfnamefont{D.}~\bibnamefont{Bouwmeester}},
  \bibinfo{journal}{Phys. Rev. Lett.} \textbf{\bibinfo{volume}{91}}, \bibinfo{pages}{130401} (\bibinfo{year}{2003}).

\bibitem[{\citenamefont{Gigan et~al.}(2006)\citenamefont{Gigan, Bohm, Paternostro, Blaser, Langer, Hertzberg, Schwab, Bauerle, Aspelmeyer, and  Zeilinger}}]{Gigan2006}
\bibinfo{author}{\bibfnamefont{S.}~\bibnamefont{Gigan}} \textit{et al.}, \bibinfo{journal}{Nature (London)} \textbf{\bibinfo{volume}{444}}, \bibinfo{pages}{67}
  (\bibinfo{year}{2006}).

\bibitem[{\citenamefont{Schliesser et~al.}(2006)\citenamefont{Schliesser,
  {Del'Haye}, Nooshi, Vahala, and Kippenberg}}]{Schliesser2006}
\bibinfo{author}{\bibfnamefont{A.}~\bibnamefont{Schliesser}} \textit{et~al.}, \bibinfo{journal}{Phys. Rev. Lett.}
  \textbf{\bibinfo{volume}{97}}, \bibinfo{pages}{243905}
  (\bibinfo{year}{2006}).

\bibitem[{\citenamefont{Teufel et~al.}(2008)\citenamefont{Teufel, Harlow,
  Regal, and Lehnert}}]{Teufel2008}
\bibinfo{author}{\bibfnamefont{J.~D.} \bibnamefont{Teufel}},
  \bibinfo{author}{\bibfnamefont{J.~W.} \bibnamefont{Harlow}},
  \bibinfo{author}{\bibfnamefont{C.~A.} \bibnamefont{Regal}}, \bibnamefont{and}
  \bibinfo{author}{\bibfnamefont{K.~W.} \bibnamefont{Lehnert}},
  \bibinfo{journal}{Phys. Rev. Lett.} \textbf{\bibinfo{volume}{101}},
  \bibinfo{pages}{197203} (\bibinfo{year}{2008}).

\bibitem[{\citenamefont{Thompson et~al.}(2008)\citenamefont{Thompson, Zwickl, Jayich, Marquardt, Girvin, and Harris}}]{Thompson2008}
\bibinfo{author}{\bibfnamefont{J.~D.} \bibnamefont{Thompson}} \textit{et al.}, \bibinfo{journal}{Nature (London)}
  \textbf{\bibinfo{volume}{452}}, \bibinfo{pages}{72} (\bibinfo{year}{2008}).

\bibitem[{\citenamefont{Rocheleau et~al.}(2009)\citenamefont{Rocheleau, Ndukum,
  Macklin, Hertzberg, Clerk, and Schwab}}]{Rocheleau2009}
\bibinfo{author}{\bibfnamefont{T.}~\bibnamefont{Rocheleau}} \textit{et~al.},
  \bibinfo{journal}{Nature (London)} \textbf{\bibinfo{volume}{463}},
  \bibinfo{pages}{72} (\bibinfo{year}{2009}).

\bibitem{Teufel2011}
\bibinfo{author}{\bibfnamefont{J.~D.} \bibnamefont{Teufel}} \textit{et~al.}, 
  \bibinfo{journal}{Nature (London)} \textbf{\bibinfo{volume}{475}}, \bibinfo{pages}{359} (\bibinfo{year}{2011}).


\bibitem[{\citenamefont{Gr\"oblacher et~al.}(2009)\citenamefont{Gr\"oblacher,
  Hammerer, Vanner, and Aspelmeyer}}]{Groblacher2009}
\bibinfo{author}{\bibfnamefont{S.}~\bibnamefont{Gr\"oblacher}},
  \bibinfo{author}{\bibfnamefont{K.}~\bibnamefont{Hammerer}},
  \bibinfo{author}{\bibfnamefont{M.~R.} \bibnamefont{Vanner}},
  \bibnamefont{and}
  \bibinfo{author}{\bibfnamefont{M.}~\bibnamefont{Aspelmeyer}},
  \bibinfo{journal}{Nature (London)} \textbf{\bibinfo{volume}{460}}, \bibinfo{pages}{724} (\bibinfo{year}{2009}).

\bibitem[{\citenamefont{Teufel et~al.}(2010)\citenamefont{Teufel, Li, Allman,
  Cicak, Sirois, Whittaker, and Simmonds}}]{Teufel2010}
\bibinfo{author}{\bibfnamefont{J.~D.} \bibnamefont{Teufel}} \textit{et~al.}, 
  \bibinfo{journal}{Nature (London)} \textbf{\bibinfo{volume}{471}}, \bibinfo{pages}{204} (\bibinfo{year}{2011}).


\bibitem[{\citenamefont{Weis et~al.}(2010)\citenamefont{Weis, Rivi{\`e}re,
  Del{\'e}glise, Gavartin, Arcizet, Schliesser, and Kippenberg}}]{Weis2010}
\bibinfo{author}{\bibfnamefont{S.}~\bibnamefont{Weis}} \textit{et~al.}, \bibinfo{journal}{Science}
  \textbf{\bibinfo{volume}{330}}, \bibinfo{pages}{1520} (\bibinfo{year}{2010}).

\bibitem[{\citenamefont{Safavi-Naeini et~al.}(2010)\citenamefont{Safavi-Naeini,
  Alegre, Chan, Eichenfield, Winger, Lin, Hill, Chang, and
  Painter}}]{Safavi2010}
\bibinfo{author}{\bibfnamefont{A.~H.} \bibnamefont{Safavi-Naeini}} \textit{et al.}, \bibinfo{journal}{Nature (London)} \textbf{\bibinfo{volume}{472}}, \bibinfo{pages}{69} (\bibinfo{year}{2011}).

\bibitem[{\citenamefont{Akram et~al.}(2010)\citenamefont{Akram, Kiesel,
  Aspelmeyer, and Milburn}}]{Akram2010}
\bibinfo{author}{\bibfnamefont{U.}~\bibnamefont{Akram}},
  \bibinfo{author}{\bibfnamefont{N.}~\bibnamefont{Kiesel}},
  \bibinfo{author}{\bibfnamefont{M.}~\bibnamefont{Aspelmeyer}},
  \bibnamefont{and} \bibinfo{author}{\bibfnamefont{G.~J.}
  \bibnamefont{Milburn}}, \bibinfo{journal}{New J. Phys.}
  \textbf{\bibinfo{volume}{12}}, \bibinfo{pages}{083030}
  (\bibinfo{year}{2010}).

\bibitem[{\citenamefont{B\o{}rkje et~al.}(2011)\citenamefont{B\o{}rkje,
  Nunnenkamp, and Girvin}}]{Borkje2011}
\bibinfo{author}{\bibfnamefont{K.}~\bibnamefont{B\o{}rkje}},
  \bibinfo{author}{\bibfnamefont{A.}~\bibnamefont{Nunnenkamp}},
  \bibnamefont{and} \bibinfo{author}{\bibfnamefont{S.~M.} \bibnamefont{Girvin}}, \bibinfo{note}{arXiv:1103.2368}.

\bibitem[{\citenamefont{{O'Connell} et~al.}(2010)\citenamefont{{O'Connell},
  Hofheinz, Ansmann, Bialczak, Lenander, Lucero, Neeley, Sank, Wang, Weides
  et~al.}}]{OConnell2010}
\bibinfo{author}{\bibfnamefont{A.~D.} \bibnamefont{{O'Connell}}}
  \textit{et~al.}, \bibinfo{journal}{Nature (London)}
  \textbf{\bibinfo{volume}{464}}, \bibinfo{pages}{697} (\bibinfo{year}{2010}).

\bibitem[{\citenamefont{Sankey et~al.}(2010)\citenamefont{Sankey, Yang, Zwickl,
  Jayich, and Harris}}]{Sankey2010}
\bibinfo{author}{\bibfnamefont{J.~C.} \bibnamefont{Sankey}} \textit{et~al.}, \bibinfo{journal}{Nature Physics}
  \textbf{\bibinfo{volume}{6}}, \bibinfo{pages}{707} (\bibinfo{year}{2010}).

\bibitem[{\citenamefont{Purdy et~al.}(2010)\citenamefont{Purdy, Brooks, Botter,
  Brahms, Ma, and Stamper-Kurn}}]{Purdy2010}
\bibinfo{author}{\bibfnamefont{T.~P.} \bibnamefont{Purdy}} \textit{et~al.},
  \bibinfo{journal}{Phys. Rev. Lett.} \textbf{\bibinfo{volume}{105}},
  \bibinfo{pages}{133602} (\bibinfo{year}{2010}).

\bibitem[{\citenamefont{Nunnenkamp et~al.}(2010)\citenamefont{Nunnenkamp,
  B\o{}rkje, Harris, and Girvin}}]{Nunnenkamp2010}
\bibinfo{author}{\bibfnamefont{A.}~\bibnamefont{Nunnenkamp}},
  \bibinfo{author}{\bibfnamefont{K.}~\bibnamefont{B\o{}rkje}},
  \bibinfo{author}{\bibfnamefont{J.~G.~E.} \bibnamefont{Harris}},
  \bibnamefont{and} \bibinfo{author}{\bibfnamefont{S.~M.}
  \bibnamefont{Girvin}}, \bibinfo{journal}{Phys. Rev. A}
  \textbf{\bibinfo{volume}{82}}, \bibinfo{pages}{021806(R)}
  (\bibinfo{year}{2010}).

\bibitem[{\citenamefont{Gupta et~al.}(2007)\citenamefont{Gupta, Moore, Murch,
  and Stamper-Kurn}}]{Gupta2007}
\bibinfo{author}{\bibfnamefont{S.}~\bibnamefont{Gupta}},
  \bibinfo{author}{\bibfnamefont{K.~L.} \bibnamefont{Moore}},
  \bibinfo{author}{\bibfnamefont{K.~W.} \bibnamefont{Murch}}, \bibnamefont{and}
  \bibinfo{author}{\bibfnamefont{D.~M.} \bibnamefont{Stamper-Kurn}},
  \bibinfo{journal}{Phys. Rev. Lett.} \textbf{\bibinfo{volume}{99}},
  \bibinfo{pages}{213601} (\bibinfo{year}{2007}).

\bibitem[{\citenamefont{Eichenfield et~al.}(2009)\citenamefont{Eichenfield,
  Chan, Camacho, Vahala, and Painter}}]{Eichenfield2009}
\bibinfo{author}{\bibfnamefont{M.}~\bibnamefont{Eichenfield}} \textit{et~al.},
  \bibinfo{journal}{Nature (London)} \textbf{\bibinfo{volume}{462}},
  \bibinfo{pages}{78} (\bibinfo{year}{2009}).

\bibitem{Mancini1997}
\bibinfo{author}{\bibfnamefont{S.}~\bibnamefont{Mancini}},
  \bibinfo{author}{\bibfnamefont{V.~I.}~\bibnamefont{Man'ko}}, \bibnamefont{and} \bibinfo{author}{\bibfnamefont{P.} \bibnamefont{Tombesi}},
  \bibinfo{journal}{Phys. Rev. A} \textbf{\bibinfo{volume}{55}},
  \bibinfo{pages}{3042} (\bibinfo{year}{1997}).

\bibitem{Bose1997}
\bibinfo{author}{\bibfnamefont{S.}~\bibnamefont{Bose}},
  \bibinfo{author}{\bibfnamefont{K.}~\bibnamefont{Jacobs}}, \bibnamefont{and} \bibinfo{author}{\bibfnamefont{P.~L.} \bibnamefont{Knight}},
  \bibinfo{journal}{Phys. Rev. A} \textbf{\bibinfo{volume}{56}},
  \bibinfo{pages}{4175} (\bibinfo{year}{1997}).

\bibitem[{\citenamefont{Ludwig et~al.}(2008)\citenamefont{Ludwig, Kubala, and Marquardt}}]{Ludwig2008}
\bibinfo{author}{\bibfnamefont{M.}~\bibnamefont{Ludwig}},
  \bibinfo{author}{\bibfnamefont{B.}~\bibnamefont{Kubala}}, \bibnamefont{and}
  \bibinfo{author}{\bibfnamefont{F.}~\bibnamefont{Marquardt}},
  \bibinfo{journal}{New J. Phys.} \textbf{\bibinfo{volume}{10}},
  \bibinfo{pages}{095013} (\bibinfo{year}{2008}).

\bibitem[{\citenamefont{Clerk et~al.}(2010)\citenamefont{Clerk, Devoret, Girvin, Marquardt, and Schoelkopf}}]{GirvinRMP}
\bibinfo{author}{\bibfnamefont{A.~A.} \bibnamefont{Clerk}} \textit{et~al.},
  \bibinfo{journal}{Rev. Mod. Phys.} \textbf{\bibinfo{volume}{82}}, \bibinfo{pages}{1155} (\bibinfo{year}{2010}).

\bibitem[{\citenamefont{Murch et~al.}(2008)\citenamefont{Murch, Moore, Gupta, and {Stamper-Kurn}}}]{Murch2008}
\bibinfo{author}{\bibfnamefont{K.~W.} \bibnamefont{Murch}},
  \bibinfo{author}{\bibfnamefont{K.~L.} \bibnamefont{Moore}},
  \bibinfo{author}{\bibfnamefont{S.}~\bibnamefont{Gupta}}, \bibnamefont{and}
  \bibinfo{author}{\bibfnamefont{D.~M.} \bibnamefont{{Stamper-Kurn}}},
  \bibinfo{journal}{Nature Physics} \textbf{\bibinfo{volume}{4}},
  \bibinfo{pages}{561} (\bibinfo{year}{2008}).

\bibitem[{\citenamefont{Mahan}(2000)}]{Mahan2000}
\bibinfo{author}{\bibfnamefont{G.~M.} \bibnamefont{Mahan}},
  \emph{\bibinfo{title}{Many-{P}article {P}hysics}} (\bibinfo{publisher}{Kluwer
  {A}cademic}, \bibinfo{address}{New {Y}ork}, \bibinfo{year}{2000}),
  \bibinfo{edition}{3rd} ed.

\bibitem[{\citenamefont{Koch et~al.}(2005)\citenamefont{Koch, Raikh, and von Oppen}}]{Koch2005}
\bibinfo{author}{\bibfnamefont{J.}~\bibnamefont{Koch}},
  \bibinfo{author}{\bibfnamefont{M.~E.} \bibnamefont{Raikh}}, \bibnamefont{and}
  \bibinfo{author}{\bibfnamefont{F.}~\bibnamefont{von Oppen}},
  \bibinfo{journal}{Phys. Rev. Lett.} \textbf{\bibinfo{volume}{95}},
  \bibinfo{pages}{056801} (\bibinfo{year}{2005}).

\bibitem[{\citenamefont{Rigo et~al.}(1997)\citenamefont{Rigo, Alber, Mota-Furtado, and O'Mahony}}]{Rigo1997}
\bibinfo{author}{\bibfnamefont{M.}~\bibnamefont{Rigo}},
  \bibinfo{author}{\bibfnamefont{G.}~\bibnamefont{Alber}},
  \bibinfo{author}{\bibfnamefont{F.}~\bibnamefont{Mota-Furtado}},
  \bibnamefont{and} \bibinfo{author}{\bibfnamefont{P.~F.}
  \bibnamefont{O'Mahony}}, \bibinfo{journal}{Phys. Rev. A}
  \textbf{\bibinfo{volume}{55}}, \bibinfo{pages}{1665} (\bibinfo{year}{1997}).
  
\bibitem{Rabl2011}
\bibinfo{author}{\bibfnamefont{P.}~\bibnamefont{Rabl}}, \bibinfo{note}{arXiv:1102:0278}.

\bibitem[{\citenamefont{Tan}(1999)}]{Tan1999}
\bibinfo{author}{\bibfnamefont{S.~M.} \bibnamefont{Tan}},
  \bibinfo{journal}{J. Opt. B} \textbf{\bibinfo{volume}{1}}, \bibinfo{pages}{424} (\bibinfo{year}{1999}).

\end{thebibliography}
\end{document}